\documentclass[runningheads]{llncs}

\usepackage[T1]{fontenc}
\usepackage{graphicx}
\usepackage{cite}
\usepackage{amsmath,amssymb}
\usepackage{booktabs}
\usepackage{array}
\usepackage{placeins}
\usepackage{longtable}
\usepackage{pdflscape}
\usepackage{hyperref}
\usepackage{xcolor}
\usepackage{tikz}
\usetikzlibrary{arrows.meta,backgrounds,fit,positioning}
\usepackage{multirow}

\begin{document}

\title{\texorpdfstring{System-Level Optimization Beyond Cryptographic Kernels:\\
An ML-KEM Case Study on Arm Cortex-M7}{System-Level Optimization Beyond
Cryptographic Kernels: An ML-KEM Case Study on Arm Cortex-M7}}

\titlerunning{System-Level Optimization Beyond Cryptographic Kernels}

\author{Mahmoud Abdelhafeez Sayed\inst{1} \and
Mostafa Taha\inst{1} \and Gurp Nijjer\inst{2}}
\authorrunning{M. A. Sayed et al.}

\institute{Department of Systems and Computer Engineering, Carleton University,
Ottawa, ON, Canada\\
\email{\{Mahmoud.Sayed3,Mostafa.Taha\}@Carleton.ca}
\and
Quantegra Technologies Inc., White Rock, BC, Canada\\
\email{gurp@quantegra.ca}}

\maketitle

\begin{abstract}\emergencystretch=1em

Recent work on embedded post-quantum cryptography has focused primarily on instruction-level optimization, including arithmetic-kernel improvements, assembly tuning, register allocation, and instruction scheduling. Using the Module-Lattice-Based Key-Encapsulation Mechanism (ML-KEM) on an Arm Cortex-M7 as a case study, we examine the additional gains available from memory-hierarchy utilization, tightly coupled memory placement, peripheral integration, clock configuration, and deterministic public-data reuse. The evaluation starts from a state-of-the-art SLOTHY-optimized implementation and covers all three ML-KEM parameter sets. Without modifying the cryptographic algorithm or standardized wire formats, the evaluated profiles without auxiliary public state reduce cycles by up to 2.5\%. A selected public-data-reuse profile reduces encapsulation and decapsulation cycles by up to 74.6\% and 58.8\%, respectively. These results demonstrate that substantial deployment gains remain after arithmetic-kernel optimization and motivate a two-stage methodology that also examines the surrounding execution system.

\end{abstract}

\noindent\textbf{Extended-version note.} This manuscript is an extended version of the corresponding accepted conference paper.

\section{Introduction}
Large-scale quantum computers would compromise the security assumptions of widely deployed public-key cryptosystems based on integer factorization and discrete logarithms~\cite{shor1994algorithms}. This threat has motivated the transition to post-quantum cryptography (PQC), culminating in the publication of the first National Institute of Standards and Technology (NIST) PQC standards in 2024~\cite{nist_pqc_standards_2024}. Among these standards, Federal Information Processing Standard (FIPS) 203 specifies the Module-Lattice-Based Key-Encapsulation Mechanism (ML-KEM), derived from CRYSTALS-Kyber, for post-quantum key establishment~\cite{nist_fips203}. ML-KEM is expected to become a central primitive in future secure communication protocols, including those deployed on embedded and Internet-of-Things (IoT) devices.

Deploying PQC on microcontrollers remains challenging. Although lattice-based schemes provide attractive performance compared with many other PQC candidates, they require substantially more computation and memory than classical public-key primitives in many embedded use cases. ML-KEM relies heavily on polynomial arithmetic, number-theoretic transforms (NTTs), SHAKE-based expansion, hashing, rejection sampling, and structured matrix-vector operations~\cite{nist_fips203}. Consequently, implementation performance depends not only on arithmetic kernels, but also on memory placement, memory hierarchy utilization, random-number generation, clock configuration, and the cost of repeatedly deriving public data.

Recent work has made substantial progress on high-speed PQC implementations for Arm microcontrollers, from the \texttt{pqm4} benchmarking framework and Cortex-M4 arithmetic kernels to optimized Keccak and SLOTHY-assisted migration of assembly implementations for ML-KEM and the Module-Lattice-Based Digital Signature Algorithm (ML-DSA) to Cortex-M7~\cite{pqm4,abdulrahman2022faster,adomnicai2023keccak,abdulrahman2025slothy}. These works focus primarily on instruction-level optimization through arithmetic improvements, assembly tuning, instruction scheduling, register allocation, and software pipelining. Such optimizations minimize the computational work required by the cryptographic algorithm.

Once these kernels have been optimized, performance can remain limited by their execution environment and by work repeated across application programming interface (API) calls.

In this paper, we propose a two-stage methodology for embedded optimization. It complements instruction-level tuning with systematic evaluation of the execution system, including memory placement, hardware integration, operating points, and deterministic public state across operation and key lifecycles. We instantiate the methodology using ML-KEM on the Arm Cortex-M7, starting from the optimized implementation of Abdulrahman et al.~\cite{abdulrahman2025slothy}. The study evaluates all three ML-KEM parameter sets and reports cycle counts, wall-clock latency, memory consumption, and caller-managed auxiliary public state while preserving the standardized public-key, secret-key, ciphertext, and shared-secret formats.

A particular motivation for the clock-scaling study is that the state-of-the-art Cortex-M7 implementation of ML-KEM~\cite{abdulrahman2025slothy} focuses on instruction-level optimization and reports its measurements at 24~MHz to eliminate Flash instruction-fetch wait states. While appropriate for evaluating computational efficiency, such measurements do not directly answer a deployment question: what is the lowest achievable execution latency on the target platform? Because cycle counts vary with memory-system behavior and are not strictly frequency invariant, we evaluate ML-KEM across the full operating range of the platform.

The contributions of this paper are as follows:

\begin{itemize}

\item We formulate a two-stage view that applies platform integration and optimization across operation lifecycles around an optimized cryptographic-kernel baseline.

\item We experimentally study Cortex-M7 system-level optimization knobs, including instruction tightly coupled memory (ITCM) placement, Flash acceleration, data tightly coupled memory (DTCM) placement of public constants, hardware true-random-number-generator (TRNG) integration, and clock configuration, under a controlled clock sweep from 24~MHz to 216~MHz.

\item We identify repeated public-data derivation as a major remaining cost in optimized ML-KEM and design wire-format-preserving public-data reuse techniques for ML-KEM-512, ML-KEM-768, and ML-KEM-1024.

\item We derive practical system-level optimization classes, selection criteria, and deployment-oriented profile-selection guidelines. The methodology transfers as a set of engineering questions, while the measured rankings and gains remain platform and workload dependent.

\end{itemize}

The results show that substantial additional gains remain possible after low-level arithmetic optimization. The evaluated profiles without auxiliary public state reduce cycles by up to 2.5\% without changing cryptographic code, while the selected public-data-reuse profile reduces encapsulation and decapsulation cycles by up to 74.6\% and 58.8\%, respectively.

The remainder of this paper is organized as follows. Section~\ref{sec:related} reviews related work on embedded post-quantum cryptographic implementations. Section~\ref{sec:framework} introduces the proposed two-stage optimization framework and derives general system-level optimization classes and selection criteria. Section~\ref{sec:optimization_directions} instantiates the framework using ML-KEM on Cortex-M7 and describes the evaluated optimization directions. Section~\ref{sec:implementation_benchmarking} presents the implementation and experimental evaluation results. Section~\ref{sec:discussion_optimum_profiles} discusses deployment-oriented profile selection and broader lessons for system-level optimization. Finally, Section~\ref{sec:conclusion} concludes the
paper.

\section{Related Work}
\label{sec:related}

Efficient embedded implementations have been an important part of the PQC evaluation process. The \texttt{pqm4} project established a common framework for testing and benchmarking NIST PQC candidates on Arm Cortex-M microcontrollers and became a widely used baseline for embedded cryptographic evaluation~\cite{pqm4}. This enabled a sequence of increasingly optimized implementations of lattice-based schemes, especially Kyber/ML-KEM and Dilithium/ML-DSA, whose dominant costs are polynomial arithmetic, NTTs, hashing, sampling, and matrix-vector operations.

Most prior work concentrates on arithmetic and assembly kernels. For Kyber on Cortex-M4, Botros et al.\ optimized NTT execution and utilized digital-signal-processing (DSP) features of the Armv7E-M instruction set to improve speed while controlling memory usage~\cite{botros2019kyber}. Alkim et al.\ further investigated improved reductions, NTT restructuring, and implementation trade-offs for ring and module variants of Learning with Errors (LWE) on Cortex-M4~\cite{alkim2020cortexm4}. Abdulrahman et al.\ later improved Kyber and Dilithium through optimized polynomial arithmetic, faster Barrett reductions, and alternative transform structures~\cite{abdulrahman2022faster}, while Huang et al.\ accelerated lattice arithmetic using improved Plantard arithmetic techniques~\cite{huang2022plantard}. Similarly, Adomnicai improved Keccak performance on Armv7-M platforms by optimizing instruction scheduling and memory-access behavior~\cite{adomnicai2023keccak}. These works primarily reduce the computational cost of the cryptographic kernels themselves.

More recently, Abdulrahman et al.\ introduced SLOTHY-based migration of highly optimized Cortex-M4 assembly implementations to Cortex-M7, extending automated instruction scheduling, register allocation, and software pipelining to the newer microarchitecture~\cite{abdulrahman2024slothy,abdulrahman2025slothy}. Their work represents the current state of the art in instruction-level optimization for ML-KEM and ML-DSA on Cortex-M7 and serves as the baseline for the present study.

In this context, \texttt{pqm4} denotes the Cortex-M4 implementation and benchmarking collection, whereas \texttt{pqm7} denotes the Cortex-M7-targeted implementation and benchmarking collection; these names are software targets rather than instruction sets. Cortex-M4 and Cortex-M7 both implement Armv7E-M and can execute the same assembly instructions, but the M7 has a dual-issue pipeline with substantially different scheduling constraints. The SLOTHY migration therefore reschedules and reallocates existing M4 kernels for the M7 microarchitecture rather than relying on M7-only instructions~\cite{pqm4,abdulrahman2025slothy}.

A smaller body of work has explored memory-performance trade-offs. For Saber, Karmakar et al.\ treated public-matrix generation as an implementation trade-off to reduce memory requirements on embedded platforms~\cite{karmakar2018saber}. Howe and Westerbaan showed that Cortex-M7 microarchitectural characteristics can significantly influence the observed performance of lattice-based PQC implementations~\cite{howe2023cortexm7}. These studies demonstrate that practical performance depends not only on arithmetic kernels but also on the interaction between the implementation and the underlying platform.

This paper is complementary to the above work. Rather than proposing new arithmetic kernels, transform structures, reductions, or assembly schedules, we start from the optimized baseline and investigate memory-hierarchy utilization, tightly coupled memory placement, public-data reuse, peripheral integration, and clock configuration. The contribution is to organize and evaluate these choices together within the two-stage framework of Section~\ref{sec:framework}.

\section{A Two-Stage View of Embedded Cryptographic Optimization}
\label{sec:framework}

We organize embedded optimization into two stages. \emph{Instruction-level optimization} reduces local computational cost through arithmetic design, transform restructuring, instruction scheduling, register allocation, software pipelining, and assembly implementation. \emph{System-level optimization} addresses the execution and reuse of those kernels within a deployment through memory placement, peripheral integration, operating-point selection, and management of deterministic state across API calls.

\begin{figure}[t]
\centering
\begin{tikzpicture}[
box/.style={
draw,
rectangle,
align=center,
minimum width=4.0cm,
minimum height=0.45cm,
inner sep=2pt
}
]

\node[box] (alg)
{Cryptographic Algorithm};

\node[box, below=4mm of alg] (inst)
{\textbf{Instruction-Level Optimization}\\
\scriptsize Arithmetic, Scheduling, Assembly};

\node[box, below=4mm of inst] (kernel)
{Optimized Kernel};

\node[box, below=4mm of kernel] (sys)
{\textbf{System-Level Optimization}\\
\scriptsize Memory, Reuse, TRNG, Clock};

\node[box, below=4mm of sys] (final)
{Optimized Embedded System};

\draw[->] (alg) -- (inst);
\draw[->] (inst) -- (kernel);
\draw[->] (kernel) -- (sys);
\draw[->] (sys) -- (final);

\end{tikzpicture}
\caption{Two-stage optimization framework.}
\label{fig:two_stage_framework}
\end{figure}
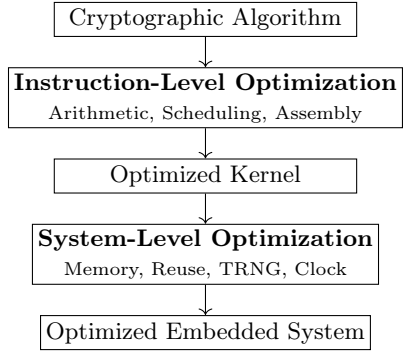

The stages identify where an optimization acts rather than imposing a strict division between algorithm-dependent and target-dependent techniques. Public-matrix reuse, for example, follows the ML-KEM data flow, whereas ITCM placement follows the processor memory system; both can be applied around an optimized kernel. Figure~\ref{fig:two_stage_framework} summarizes this view.

\subsection{System-Level Optimization Classes}
\label{subsec:optimization_classes}

The optimization opportunities encountered in embedded cryptographic workloads can be grouped into a small number of recurring system-level optimization classes:

\begin{itemize}
\item \textbf{Hot-code placement} places frequently executed code regions into low-latency instruction memories to reduce instruction-fetch overhead.
\item \textbf{Hot-data placement} relocates frequently accessed data structures into fast data memories to reduce access latency and improve execution predictability.
\item \textbf{Public-data reuse} trades additional storage for reduced recomputation by caching deterministic values derived exclusively from public information.
\item \textbf{Hardware offloading} utilizes platform peripherals or accelerators to reduce software overhead and free processor resources.
\item \textbf{Clock and operating-point optimization} adjusts processor and memory operating parameters to reduce wall-clock execution time or improve energy efficiency.
\end{itemize}

\subsection{Optimization Selection Criteria}
\label{subsec:selection_rules}

The classes above are engineering categories rather than predictive performance models. Candidate selection should combine static properties, such as code or data size, with measured execution frequency, recomputation cost, and platform behavior. Table~\ref{tab:selection_criteria} gives a concise screening view; the discussion below explains how each criterion should be interpreted.

\begin{table}[t]
\centering
\caption{Practical criteria for selecting system-level optimizations.}
\label{tab:selection_criteria}
\small
\setlength{\tabcolsep}{3pt}
\begin{tabular}{>{\raggedright\arraybackslash}p{0.18\textwidth}>{\raggedright\arraybackslash}p{0.34\textwidth}>{\raggedright\arraybackslash}p{0.34\textwidth}}
\toprule
Class & Favorable evidence & Principal cost or constraint \\
\midrule
Hot code & Frequent execution and measured fetch pressure relative to code size
& Limited fast-code capacity and active-path placement interactions \\
Hot data & Frequent loads and reuse with measurable data-access pressure
& RAM footprint, startup copying, and competition for fast data memory \\
Public reuse & Expensive deterministic derivation repeated under a stable key association
& Preparation, storage, distribution, cache lifetime, and key binding \\
Hardware & Software cost exceeds peripheral setup and service cost
& Availability, driver latency, integration complexity, and security requirements \\
Operating point & Lower measured latency under validated memory and peripheral settings
& Power, energy, thermal limits, wait states, and peripheral timing \\
\bottomrule
\end{tabular}
\end{table}
\FloatBarrier

\paragraph{Hot-code placement.}
The strongest candidates are compact routines that execute frequently and exhibit instruction-fetch pressure. Execution frequency alone is insufficient: moving a large routine may displace several smaller routines whose aggregate benefit is greater, and code that is hot in one execution path may be inactive in another. Selection should therefore combine call or cycle profiles, symbol sizes, fast-memory capacity, and measurements of candidate groups. Placement must be reevaluated when the parameter set, clock, or algorithmic path changes.

\paragraph{Hot-data placement.}
Frequently loaded and repeatedly reused tables or state are natural candidates for low-latency data memory, but only when their accesses contribute materially to execution time. Data already served effectively by a cache may obtain little benefit from relocation. The evaluation must also include the static RAM consumed, any startup copy from nonvolatile storage, alignment requirements, and the opportunity cost of excluding other data from the same fast-memory region.

\paragraph{Public-data reuse.}
Reuse is attractive when deterministic public information is expensive to derive and is consumed repeatedly while its association with a key remains stable. The relevant comparison is lifecycle based: preparation, storage, and possible distribution costs must be weighed against the accumulated recomputation avoided over later operations. Because the state is public, confidentiality is not the primary concern, but correct key binding, authenticity, integrity, invalidation, and the duration for which the cache is retained remain part of the design.

\paragraph{Hardware offloading.}
A peripheral is useful when the processor work it replaces exceeds setup, synchronization, data-transfer, and driver costs. Availability alone does not imply a speedup, and an offloaded path must still satisfy the cryptographic function's security requirements. For randomness, for example, entropy quality, health testing, error handling, and service latency matter in addition to cycle count. Measurements should include the integration overhead observed by the calling cryptographic operation.

\paragraph{Operating-point selection.}
Applications observe latency rather than clock frequency in isolation. A useful operating point must therefore be validated with the actual clock tree, Flash wait states, caches, memories, and peripheral timings used by the deployment. Higher frequency can reduce latency even if the measured cycle count rises slightly, but it can also increase power or thermal demand. Energy or battery lifetime cannot be inferred from timing alone and requires separate power measurement.

\subsection{Scope and Transferability}
\label{subsec:scope_transferability}

The classes and criteria above transfer as questions to ask of other embedded cryptographic workloads, but the measured gains do not transfer unchanged. A hot-code strategy may use ITCM on Cortex-M7, SRAM on a lower-end Cortex-M device, or a scratchpad on an embedded RISC-V processor; the available capacity and access behavior differ in each case. Algorithms also differ in their hardware interfaces and reusable public state.

The Cortex-M7 results in this paper therefore demonstrate the methodology on one platform rather than establish a universal ranking of optimization classes. The remainder of the paper instantiates the criteria using ML-KEM and measures the resulting choices on the target system.
\FloatBarrier

\section{ML-KEM System-Level Optimization on Cortex-M7}
\label{sec:optimization_directions}

We apply the classes of Section~\ref{sec:framework} to the SLOTHY-optimized Cortex-M7 baseline~\cite{abdulrahman2025slothy}.

ML-KEM key generation produces a public/secret key pair. Encapsulation uses the public key and fresh randomness to produce a ciphertext and shared secret; decapsulation uses the secret key to recover the corresponding shared secret~\cite{nist_fips203}. Internally, these operations use polynomial arithmetic accelerated by the NTT. The public key contains a polynomial vector and a seed $\rho$, expanded by SHAKE and rejection sampling into a public matrix $A$ in NTT form. Key generation multiplies $A$ by a secret vector and adds noise; encapsulation uses $A^T$ and a dot product with the public-key vector. Decapsulation decrypts and re-encrypts a candidate message to check the ciphertext, repeating the encryption arithmetic and returning a fallback secret on mismatch.

In all cases, the standard ML-KEM public-key, secret-key, ciphertext, and shared-secret formats are preserved; any additional state is caller-managed implementation storage.

\subsection{Hot-Code Placement}
\label{subsec:hot_code_placement}

The hot-code-placement class corresponds to optimizations that improve instruction fetch behavior by relocating frequently executed code regions into low-latency instruction memory. In the experimental evaluation of Section~\ref{sec:implementation_benchmarking}, this optimization class is represented by \emph{Endpoint-selected ITCM} and \emph{Flash acceleration}.

\subsubsection{ITCM Placement of Hot Code}
\label{subsubsec:itcm}

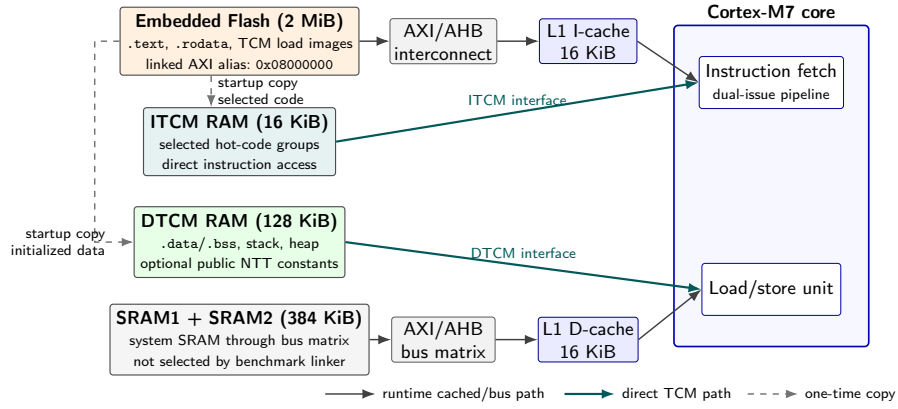
\begin{figure}[!htbp]
\centering
\resizebox{0.98\textwidth}{!}{\begin{tikzpicture}[
  font=\sffamily\footnotesize,
  >=Latex,
  memory/.style={
    draw=black!70,
    rounded corners=1.5pt,
    line width=0.55pt,
    minimum width=3.25cm,
    minimum height=1.05cm,
    align=center,
    inner sep=3pt
  },
  interconnect/.style={
    draw=black!65,
    rounded corners=1.5pt,
    fill=black!5,
    minimum width=1.75cm,
    minimum height=0.72cm,
    align=center,
    inner sep=2pt
  },
  cache/.style={
    draw=blue!55!black,
    rounded corners=1.5pt,
    fill=blue!8,
    minimum width=1.65cm,
    minimum height=0.72cm,
    align=center,
    inner sep=2pt
  },
  coreunit/.style={
    draw=blue!65!black,
    rounded corners=1.5pt,
    fill=white,
    minimum width=2.35cm,
    minimum height=0.88cm,
    align=center,
    inner sep=3pt
  },
  runtime/.style={-{Latex[length=2.2mm]}, line width=0.7pt, draw=black!75},
  tcmpath/.style={-{Latex[length=2.2mm]}, line width=1.05pt, draw=teal!70!black},
  startup/.style={-{Latex[length=2mm]}, line width=0.7pt, dashed, draw=black!55}
]

\node[memory, fill=orange!12] (flash) at (1.75,5.25)
  {\textbf{Embedded Flash (2 MiB)}\\[-1pt]
   \scriptsize \texttt{.text}, \texttt{.rodata}, TCM load images\\[-1pt]
   \scriptsize linked AXI alias: \texttt{0x08000000}};

\node[memory, fill=teal!10] (itcm) at (1.75,3.55)
  {\textbf{ITCM RAM (16 KiB)}\\[-1pt]
   \scriptsize selected hot-code groups\\[-1pt]
   \scriptsize direct instruction access};

\node[memory, fill=green!10] (dtcm) at (1.75,1.85)
  {\textbf{DTCM RAM (128 KiB)}\\[-1pt]
   \scriptsize \texttt{.data}/\texttt{.bss}, stack, heap\\[-1pt]
   \scriptsize optional public NTT constants};

\node[memory, fill=black!4] (sram) at (1.75,0.20)
  {\textbf{SRAM1 + SRAM2 (384 KiB)}\\[-1pt]
   \scriptsize system SRAM through bus matrix\\[-1pt]
   \scriptsize not selected by benchmark linker};

\node[interconnect] (axi_i) at (5.20,5.25) {AXI/AHB\\interconnect};
\node[cache] (icache) at (7.65,5.25) {L1 I-cache\\16 KiB};
\node[interconnect] (axi_d) at (5.20,0.20) {AXI/AHB\\bus matrix};
\node[cache] (dcache) at (7.65,0.20) {L1 D-cache\\16 KiB};

\node[coreunit] (fetch) at (10.75,4.55)
  {Instruction fetch\\\scriptsize dual-issue pipeline};
\node[coreunit] (loadstore) at (10.75,1.05)
  {Load/store unit};
\begin{scope}[on background layer]
\node[
  draw=blue!70!black,
  rounded corners=2pt,
  line width=0.8pt,
  fill=blue!3,
  fit=(fetch)(loadstore),
  inner xsep=0.42cm,
  inner ysep=0.52cm,
  label={[font=\sffamily\bfseries\small]above:Cortex-M7 core}
] (core) {};
\end{scope}

\draw[runtime] (flash.east) -- (axi_i.west);
\draw[runtime] (axi_i.east) -- (icache.west);
\draw[runtime] (icache.east) -- (fetch.west);

\draw[tcmpath] (itcm.east) -- node[above, font=\sffamily\scriptsize,
  text=teal!70!black] {ITCM interface} (fetch.west);
\draw[tcmpath] (dtcm.east) -- node[above, font=\sffamily\scriptsize,
  text=teal!70!black] {DTCM interface} (loadstore.west);

\draw[runtime] (sram.east) -- (axi_d.west);
\draw[runtime] (axi_d.east) -- (dcache.west);
\draw[runtime] (dcache.east) -- (loadstore.west);

\draw[startup] ([xshift=-0.48cm]flash.south) --
  node[right, font=\sffamily\scriptsize, align=left] {startup copy\\selected code}
  ([xshift=-0.48cm]itcm.north);
\draw[startup] (flash.west) -- ++(-0.42,0) |-
  node[pos=0.73, left, font=\sffamily\scriptsize, align=right]
  {startup copy\\initialized data}
  (dtcm.west);

\draw[runtime] (3.20,-0.72) -- (4.05,-0.72)
  node[right, font=\sffamily\scriptsize] {runtime cached/bus path};
\draw[tcmpath] (7.25,-0.72) -- (8.10,-0.72)
  node[right, font=\sffamily\scriptsize] {direct TCM path};
\draw[startup] (10.35,-0.72) -- (11.20,-0.72)
  node[right, font=\sffamily\scriptsize] {one-time copy};

\end{tikzpicture}}
\caption{Configured STM32F767ZI memory paths used in this study. Solid arrows denote runtime instruction or data access, teal arrows denote direct TCM access, and dashed arrows denote one-time startup copies from Flash load images. Normal code executes from the linked AXI Flash alias through the L1 instruction cache; selected hot-code groups execute from ITCM RAM. SRAM1/2 are shown for architectural context but are not selected by the benchmark linker.}
\label{fig:stm32f767_memory_paths}
\end{figure}

The Cortex-M7 ITCM can serve instruction fetches with lower and more predictable latency than Flash. Figure~\ref{fig:stm32f767_memory_paths} summarizes the memory paths relevant to this study and the placement implemented by the benchmark firmware~\cite{st_an4667,st_stm32f767_datasheet}.

The STM32F767ZI provides 16~KiB of ITCM RAM. Normal \texttt{.text} and \texttt{.rodata} remain in the AXI-mapped Flash region. When an ITCM group is enabled, its Flash load image is copied to \texttt{.itcm\_text} before benchmarking and the instruction cache is invalidated; all unselected routines remain on the normal Flash path.

The main candidates include the Keccak-$f[1600]$ permutation underlying SHAKE, NTT and inverse-NTT routines, and matrix-accumulation (\texttt{matacc}) kernels, which compute a matrix-row/vector product as a sum of polynomial products. Separate dot-product kernels evaluate analogous sums for polynomial vectors, including byte-encoded inputs. We also evaluate kernels that capture $A$ during key generation or consume its cached entries later, and smaller arithmetic or packing helpers.

Because ITCM capacity is limited, placement must be selective. The useful set depends on the parameter set, operating frequency, and whether the active path generates $A$ or consumes its cached representation. We therefore select candidate groups empirically for the measured endpoint configurations rather than assume one placement is optimal throughout. For example, the ML-KEM-768 regenerated-$A$ profile in Table~\ref{tab:itcm_endpoint_selection} places the Keccak, matrix-accumulation, and dot-product groups in ITCM; the cryptographic computation and remaining code are unchanged.

\subsubsection{Flash Acceleration and Prefetch Configuration}
\label{subsubsec:flash_art}

STM32F7 devices include Flash acceleration mechanisms such as the Adaptive Real-Time (ART) accelerator and prefetch buffer. These features are intended to reduce the cost of fetching instructions from Flash, particularly when the processor clock is high relative to Flash access latency.

This direction evaluates whether explicitly enabling Flash acceleration improves full ML-KEM operation timing under the selected benchmark configuration. For cold-code execution or workloads with poor instruction-cache locality, Flash acceleration may be useful. For a warm-cache benchmark dominated by tight optimized loops, the incremental benefit may be small because the instruction cache and, where enabled, ITCM already cover much of the hot path.

\subsection{Hot-Data Placement}
\label{subsec:hot_data_placement}

The hot-data-placement class corresponds to relocating frequently accessed data structures into low-latency data memory. For ML-KEM on Cortex-M7, this class is evaluated as \emph{DTCM NTT constants}.

\subsubsection{DTCM Placement of Public NTT Constants}
\label{subsubsec:dtcm_constants}

The NTT routines use public twiddle-factor tables. These constants are normally stored in Flash and accessed during polynomial arithmetic.

The DTCM-constant direction copies selected public NTT tables from Flash into DTCM at startup and directs the NTT routines to the DTCM symbols. This removes those loads from the Flash path. Since the tables are public, relocation does not introduce secret-data exposure. Its costs are startup copying and additional random-access memory (RAM), which can be significant if the measured access-time benefit is small.

\subsection{Hardware Offloading}
\label{subsec:hardware_offloading}

The hardware-offloading class utilizes platform peripherals or accelerators to reduce software overhead. For ML-KEM on STM32F7, this class is evaluated as \emph{Hardware TRNG}.

\subsubsection{Hardware TRNG Integration}
\label{subsubsec:trng}

ML-KEM key generation and encapsulation require fresh randomness. The benchmark baseline uses a deterministic software fallback, which is useful for reproducible testing but is not representative of a deployment-quality randomness source.

The hardware-TRNG direction uses the STM32 peripheral as the source for \texttt{randombytes}; the measured operations include its integration cost. This experiment compares the two paths implemented in the artifact; it does not imply that a hardware TRNG is universally faster than a deployment-quality software random-number generator. Entropy-source validation and peripheral error handling remain deployment responsibilities.

\subsection{Clock and Operating-Point Optimization}
\label{subsec:clock_operating_point}

Clock and operating-point optimization adjusts processor operating parameters to reduce wall-clock execution time without changing the underlying cryptographic computation. For Cortex-M7, this class is evaluated through the clock-configuration experiments.

\subsubsection{Clock Configuration}
\label{subsubsec:clock_scaling}
Cycle counts are the conventional metric for microcontroller cryptographic benchmarking, but wall-clock latency is the quantity directly observed by applications. The Cortex-M7 platform therefore also admits clock-configuration experiments.

Changing the processor clock does not directly reduce the number of executed instructions and does not constitute an algorithmic optimization, but it changes the latency corresponding to a given cycle count and can interact with Flash wait states, cache behavior, and peripheral timing. Consequently, clock selection is treated as an operating-point optimization within the proposed framework.

We evaluate clock profiles as a platform-level dimension of the implementation. The expected behavior is that wall-clock latency decreases approximately with increasing processor-clock frequency (HCLK) when cycle counts remain stable, while cycle counts themselves may vary slightly because memory and peripheral behavior are not perfectly invariant across clock configurations.

\subsection{Public-Data Reuse}
\label{subsec:public_data_reuse}

ML-KEM contains several computations over values that are public or derived solely from public inputs. The baseline implementation recomputes or reparses some of these values when they are needed by key generation, encapsulation, or the re-encryption step inside decapsulation.

Public-data reuse trades additional RAM and possible communication for reduced repeated work. This trade-off is particularly relevant for embedded systems in which a key is reused for many encapsulations or decapsulations. In the experimental evaluation, this optimization class is represented by \emph{Cached $A$}, \emph{Cached $H(pk)$}, where $pk$ denotes the standardized public key, and the \emph{parsed public-key cache}.

A decapsulating key owner can retain cache state prepared during key generation. A remote encapsulator must instead receive the corresponding auxiliary state, derive it once from the standardized public key, or retain a previously prepared copy. The additional bytes are therefore caller-managed auxiliary public state, not an enlargement of the standardized ML-KEM public key.

All three caches contain only public or public-key-derived values, so retaining them does not disclose an ML-KEM secret. This observation addresses confidentiality only: implementations must bind each cache to the correct public key, reject stale or mismatched state, and preserve the authenticity and integrity association of externally supplied cache data. The public origin of a cache does not by itself establish constant-time behavior or side-channel resistance.

\subsubsection{Cached Public Matrix ($A$)}
\label{subsubsec:cached_a}

The baseline regenerates $A$ whenever needed. The cached-$A$ direction instead captures a serialized, multiplication-ready representation while key generation produces the matrix and reuses the matching entries during later operations associated with the same keypair. The measured key-generation cost therefore includes materializing the caller-owned cache.

The expected benefit is largest for encapsulation and decapsulation, where repeated matrix generation can be avoided. The trade-off is additional RAM, possible distribution cost, and key-generation overhead for materializing the reusable representation.

\subsubsection{Cached $H(pk)$}
\label{subsubsec:cached_hpk}

ML-KEM encapsulation hashes the public key as part of the input to the key-derivation computation. The same hash value, $H(pk)$, is already stored in the standard ML-KEM secret key for use during decapsulation.

The cached-$H(pk)$ direction exposes this public hash as caller-managed public cache state, allowing encapsulation to reuse it instead of hashing the public key again. This direction is expected to mainly affect encapsulation. Key generation requires little additional work because the value is already available after keypair generation, and decapsulation is not expected to improve from this cache alone because the baseline decapsulation path already obtains $H(pk)$ from the secret key.

\subsubsection{Parsed Public-Key Cache}
\label{subsubsec:parsed_pk_cache}

Encapsulation and decapsulation re-encryption both use the public-key polynomial vector in ML-KEM's internal public-key encryption routine. In a conventional implementation, the byte-encoded public key is decoded and prepared when the operation is executed.

The parsed public-key cache stores this decoded public representation, and, when appropriate for the optimized backend, also stores a transformed helper representation consumed by the multiplication routines. This direction avoids repeated parsing and preparation of public-key material.

Its expected effect is smaller than cached $A$, because it does not remove public matrix generation, but it can still reduce overhead in encapsulation and in the re-encryption path of decapsulation. The storage contains only public-key-derived data and does not change the external key or ciphertext format.

\section{Implementation and Evaluation}
\label{sec:implementation_benchmarking}

This section evaluates the system-level optimization directions presented in Section~\ref{sec:optimization_directions}. The evaluation comprises 384 benchmark cases: a 48-case matrix isolating eight configurations across three parameter sets and the two endpoint clocks, and a 336-case matrix containing the baseline plus six composite profiles across three parameter sets and 16 clocks. The implementation and benchmarking harness are available in the accompanying artifact repository~\cite{MLKEMM7Repo2026}. Unless otherwise stated, a reported percentage change is $100\,(c_{\mathrm{base}}-c_{\mathrm{profile}})/c_{\mathrm{base}}$ against the baseline with the same ML-KEM parameter set and clock frequency, so positive values denote cycle reductions and negative values denote cycle increases, that is, a configuration slower than the baseline.

\subsection{Experimental Platform and Baseline}
\label{subsec:experimental_platform}

All experiments use a NUCLEO-F767ZI board with an STM32F767ZI Arm Cortex-M7. The baseline is the strongest Cortex-M7 implementation available in the repository, namely the \texttt{pqm7} backend, \texttt{m7fspeed} variant, and \texttt{m4\_opt\_m7} Keccak implementation derived from the SLOTHY-optimized Cortex-M7 path~\cite{abdulrahman2025slothy}. All additional optimization knobs are disabled in the baseline.

We evaluate the three FIPS~203 parameter sets: ML-KEM-512, ML-KEM-768, and ML-KEM-1024~\cite{nist_fips203}. Firmware is built in Release mode with the Arm GNU Toolchain and GNU Compiler Collection (GCC)~13.2.1, \texttt{-O3}, link-time optimization, and Cortex-M7 hard-float target flags. Before timed measurements, the firmware performs one warm-up call for each operation and suspends the system tick. Each operation is measured over 100 timed iterations using the Arm Data Watchpoint and Trace (DWT) cycle counter. Each measurement records minimum, arithmetic mean, median, and maximum cycles; the analysis uses the arithmetic mean. All runs report matching encapsulated and decapsulated shared secrets. Measurements include the 24 and 216~MHz endpoints, and the composite evaluation covers 16 validated clock profiles between them.

Section~\ref{subsec:measurement_repeatability} examines within-run dispersion and the repeatability of the small effects reported below.

All optimized variants preserve the standardized ML-KEM public-key, secret-key, ciphertext, and shared-secret formats. Wall-clock values are derived from mean cycles and the measured HCLK.\@ The data-RAM values are static allocations reported by the linker for the configured 128~KiB DTCM region, not peak stack measurements; they occupy 5.5--17.1\% of its capacity across the evaluated configurations. The auxiliary-state columns identify the public-cache portion of this allocation; these bytes do not alter the standardized public-key size.

Table~\ref{tab:baseline} reports the absolute baseline measurements used throughout the remainder of the paper. The results show that cycle counts remain relatively stable across clock profiles, whereas wall-clock latency scales approximately with processor frequency. This motivates reporting both metrics: cycle counts isolate implementation work, while wall-clock time captures deployment performance.

Importantly, cycle counts are not strictly frequency invariant. Across the nine baseline measurements shown in Table~\ref{tab:baseline}, the increase from 24~MHz to 216~MHz averages approximately 4.1k cycles and exceeds 5.2k cycles in the worst case. Although these increases are small relative to total execution cost, they demonstrate that execution cycles depend to some extent on operating frequency and memory-system behavior. Consequently, execution latency at higher operating frequencies cannot be inferred accurately by simply scaling low-frequency cycle measurements.

\begin{table*}[t]
\centering
\caption{Baseline cycle counts and derived execution times. Times are shown in ms in parentheses; data-RAM and standardized public-key (PK) sizes are in bytes.}
\label{tab:baseline}
\setlength{\tabcolsep}{4pt}
\scriptsize
\begin{tabular}{llrrrrr}
\toprule
Set & MHz & KeyGen & Encaps & Decaps & Data RAM & PK \\
\midrule

\multirow{2}{*}{ML-KEM-512}
& 24  & 229021 (9.54) & 229131 (9.55) & 246613 (10.28) & \multirow{2}{*}{7320} & \multirow{2}{*}{800} \\
& 216 & 232267 (1.08) & 231997 (1.07) & 250388 (1.16) & & \\
\midrule

\multirow{2}{*}{ML-KEM-768}
& 24  & 373931 (15.58) & 382991 (15.96) & 407582 (16.98) & \multirow{2}{*}{8792} & \multirow{2}{*}{1184} \\
& 216 & 378284 (1.75) & 387205 (1.79) & 412828 (1.91) & & \\
\midrule

\multirow{2}{*}{ML-KEM-1024}
& 24  & 590538 (24.61) & 597755 (24.91) & 631860 (26.33) & \multirow{2}{*}{10424} & \multirow{2}{*}{1568} \\
& 216 & 595065 (2.76) & 601529 (2.79) & 637113 (2.95) & & \\
\bottomrule
\end{tabular}
\end{table*}

\subsection{Evaluation of Individual Optimization Directions}
\label{subsec:isolated_effects}

To isolate the contribution of each system-level optimization class, we first enable one optimization direction at a time while keeping all remaining directions at their baseline values. Table~\ref{tab:isolated_effects} summarizes the resulting performance ranges across all parameter sets and the two edge-clock configurations.

\begin{table*}[t]
\centering
\caption{Isolated optimization effects relative to the baseline. Positive values denote cycle reductions (\%) and negative values denote cycle increases; data-RAM and auxiliary-public-state deltas are in bytes.}
\label{tab:isolated_effects}
\scriptsize
\setlength{\tabcolsep}{3pt}
\resizebox{\textwidth}{!}{%
\begin{tabular}{lccccc}
\toprule
Optimization & KeyGen & Encaps & Decaps & Data RAM $\Delta$ & Aux. state $\Delta$ \\
\midrule

Hardware TRNG
& 0.17 to 1.72
& 0.06 to 0.85
& -0.32 to 0.13
& -112
& 0 \\

Flash ART/prefetch
& -0.06 to -0.00
& -0.06 to 0.05
& -0.07 to 0.00
& 0
& 0 \\

Endpoint-selected ITCM
& 0.18 to 1.64
& 0.15 to 1.71
& 0.27 to 1.72
& 0
& 0 \\

DTCM NTT constants
& -0.07 to 0.00
& -0.05 to 0.05
& -0.06 to -0.01
& +1792
& 0 \\

Cached $A$
& -5.23 to -2.48
& 41.04 to 61.69
& 37.76 to 58.41
& +1536 to +6144
& +1536 to +6144 \\

$H(pk)$ cache
& -0.09 to 0.09
& 12.49 to 16.35
& -0.11 to 0.33
& +32
& +32 \\

Parsed PK cache
& -2.24 to -1.67
& 0.70 to 1.30
& 0.58 to 0.87
& +2048 to +4096
& +2048 to +4096 \\

\bottomrule
\end{tabular}%
}
\end{table*}

For each parameter set, endpoint, and profile, Table~\ref{tab:itcm_endpoint_selection} reports the measured ITCM overlay minimizing the sum of average key-generation, encapsulation, and decapsulation cycles.

\begin{table*}[t]
\centering
\caption{Endpoint-selected ITCM placements for the isolated regenerated-$A$ path and the six composite profiles.}
\label{tab:itcm_endpoint_selection}
\scriptsize
\setlength{\tabcolsep}{2pt}
\resizebox{\textwidth}{!}{%
\begin{tabular}{llccccccc}
\toprule
Set & MHz & Iso. regen-$A$ & NC & NC-noD & AC & AC-noPK & AE-noD-noPK & AE \\
\midrule
\multirow{2}{*}{ML-KEM-512}
& 24  & K+NTT & K+NTT & K+MA+DP & K+CMA & K+CMA & K+CMA & K+CMA \\
& 216 & K+NTT & K+NTT & K+NTT & K+CMA & K+NTT+CMA & K+CMA & K+CMA \\
\midrule
\multirow{2}{*}{ML-KEM-768}
& 24  & K+MA+DP & K+MA+DP & K+MA+DP & K+CMA & K+CMA & K+CMA & K+CMA \\
& 216 & K+MA+DP & K+MA+DP & K+MA+DP & K+CMA & K+NTT+CMA & K+CMA & K+NTT+CMA \\
\midrule
\multirow{2}{*}{ML-KEM-1024}
& 24  & K+MA & K+NTT & K+MA & K+CMA & K+CMA & K+CMA & K+CMA \\
& 216 & K+MA & K+MA & K+MA & K+CMA & K+CMA & K+CMA & K+CMA \\
\bottomrule
\end{tabular}%
}

\vspace{1mm}
\raggedright
\footnotesize
NC = hardware TRNG, Flash ART/prefetch, DTCM NTT constants, and selected ITCM placement; NC-noD = NC without DTCM constants; AC = cached $A$, cached $H(pk)$, and parsed-PK state; AC-noPK = AC without parsed-PK state; AE = AC plus NC; AE-noD-noPK = AE without DTCM constants or parsed-PK state.\\
For intermediate sweep clocks, 24--120~MHz use the 24~MHz selection and 144--216~MHz use the 216~MHz selection; complete candidate measurements are available in the artifact repository~\cite{MLKEMM7Repo2026}.\\
K = Keccak permutation; NTT = forward/inverse transforms; MA = matrix-row accumulation; DP = polynomial-vector dot products; CMA = matrix accumulation with $A$ capture during key generation.
\end{table*}

\begin{figure*}[t]
\centering
\includegraphics[width=\textwidth]{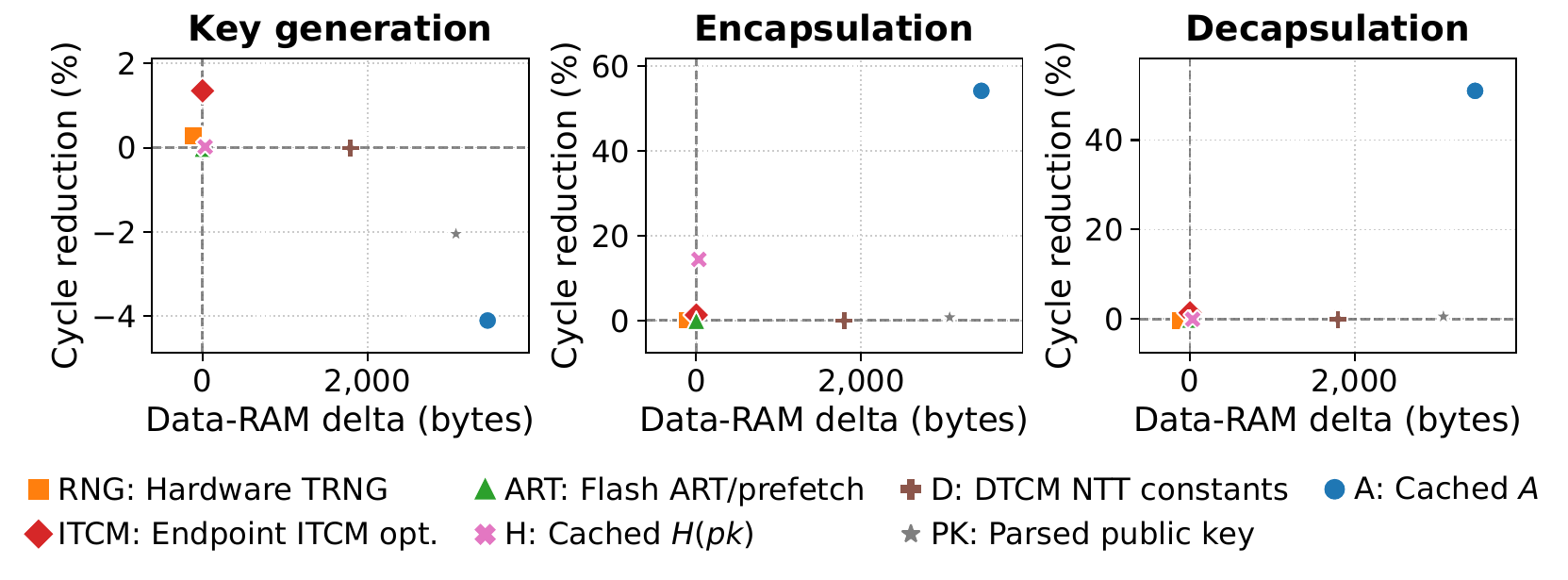}
\caption{Single-knob memory/performance trade-off for ML-KEM-768 at 216~MHz. Positive values denote cycle reductions relative to the baseline; negative values denote cycle increases.}
\label{fig:single_knob_memory_tradeoff}
\end{figure*}

Figure~\ref{fig:single_knob_memory_tradeoff} illustrates the memory-performance trade-off for a representative configuration, ML-KEM-768 at 216~MHz. Each point corresponds to a single optimization knob and shows the relationship between additional memory cost and achieved performance improvement.

The isolated measurements expose markedly different trade-offs. Cached $A$ reduces encapsulation cycles by 41.04--61.69\% and decapsulation cycles by 37.76--58.41\%, but increases key-generation cycles by 2.48--5.23\% and requires 1.5--6~KiB of auxiliary public state. Cached $H(pk)$ reduces encapsulation cycles by 12.49--16.35\% for only 32~B. By comparison, the parsed public-key cache saves at most 1.30\% while requiring 2--4~KiB.

Endpoint-selected ITCM placement is the most consistent direction that requires no auxiliary public state, reducing cycles for all three operations by 0.15--1.72\%. Hardware TRNG effects are small and include a slight decapsulation regression in some cases. Flash ART/prefetch is effectively neutral under the evaluated warm-cache protocol, while moving the public NTT constants to DTCM consumes 1792~B without a corresponding cycle benefit.

Thus, public-data reuse supplies the largest isolated gains, followed by selective hot-code placement. The results also show why available knobs should not be enabled indiscriminately: their costs and benefits must be evaluated both individually and in the complete execution path.

\subsection{Evaluation of Combined Optimization Profiles}
\label{subsec:combined_effects}

Individual optimizations provide useful insight into the contribution of each optimization class; however, practical implementations typically combine multiple optimizations simultaneously. We therefore evaluate several combined profiles to determine whether isolated gains compose effectively and to identify suitable deployment configurations.

Table~\ref{tab:combined_profiles} summarizes the six profiles across all parameter sets and both endpoint clocks. Their definitions and endpoint-specific ITCM placements are given above with Table~\ref{tab:itcm_endpoint_selection}. In these names, caching refers only to caller-managed reuse of public or public-key-derived state, not the Cortex-M7 instruction and data caches; Flash ART/prefetch is a separate hardware memory-system mechanism. Cycle reductions use the corresponding fixed-frequency baseline.

\begin{table*}[t]
\centering
\caption{Combined optimization profiles relative to the baseline. Positive values denote cycle reductions (\%) and negative values denote cycle increases; data-RAM and auxiliary-public-state deltas are in bytes.}
\label{tab:combined_profiles}
\scriptsize
\setlength{\tabcolsep}{2pt}
\begin{tabular}{lccccc}
\toprule
Profile & KeyGen & Encaps & Decaps & Data RAM $\Delta$ & Aux. state $\Delta$ \\
\midrule
NC
& 0.65 to 2.46
& 0.25 to 2.26
& 0.05 to 2.23
& +1680
& 0 \\

NC-noD
& 0.74 to 2.25
& 0.41 to 2.09
& 0.18 to 1.97
& -112
& 0 \\

AC
& -6.27 to -3.71
& 58.20 to 74.80
& 38.98 to 59.40
& +3616 to +10272
& +3616 to +10272 \\

AC-noPK
& -4.77 to -1.80
& 57.54 to 74.31
& 38.32 to 58.88
& +1568 to +6176
& +1568 to +6176 \\

AE-noD-noPK
& -4.22 to -1.08
& 58.58 to 74.60
& 38.48 to 58.78
& +1456 to +6064
& +1568 to +6176 \\

AE
& -5.66 to -3.04
& 58.96 to 75.12
& 39.04 to 59.37
& +5296 to +11952
& +3616 to +10272 \\
\bottomrule
\end{tabular}

\vspace{1mm}
\raggedright
\footnotesize
NC = system-level knobs without public-data reuse;
NC-noD = NC, no DTCM;
AC = all public-data-reuse caches;
AC-noPK = AC, no parsed public-key cache;
AE-noD-noPK = all enabled, no DTCM, no parsed public-key cache;
AE = all enabled.
\end{table*}

The profiles without public-data reuse reduce cycles by at most approximately 2.5\% and add no auxiliary public state. Public-data-reuse profiles provide much larger encapsulation and decapsulation reductions, accompanied by key-generation and storage costs. The fully enabled profile attains the largest maximum encapsulation reduction, but its DTCM constants and parsed public-key representation contribute substantial memory despite providing little benefit in isolation.

We therefore select \emph{AE-noD-noPK} as the balanced performance profile. It retains cached $A$ and $H(pk)$, hardware TRNG, Flash configuration, and path-specific ITCM placement while omitting DTCM constants and parsed public-key state. Table~\ref{tab:recommended_profile} reports its absolute results. This choice assumes that the required auxiliary public state can be retained or made available to the operation using it; it is not universally optimal.

\begin{table*}[t]
\centering
\caption{Selected all-enabled, no-DTCM, no-parsed-PK profile. Entries show
$\Delta$ (\%) compared with baseline, where positive values denote cycle reductions and negative values denote cycle increases, average cycles, and time (ms) in parentheses. Data-RAM and auxiliary-public-state values are reported in bytes.}
\label{tab:recommended_profile}
\scriptsize
\setlength{\tabcolsep}{3pt}
\resizebox{\textwidth}{!}{%
\begin{tabular}{llrrrrr}
\toprule
Set & MHz & KeyGen & Encaps & Decaps & Data RAM & Aux. state \\
\midrule

\multirow{2}{*}{ML-KEM-512}
& 24  & -1.08, 231533 (9.65) & 58.58, 94948 (3.96) & 38.48, 151693 (6.32) &
\multirow{2}{*}{8776} & \multirow{2}{*}{+1568} \\

& 216 & -1.09, 234723 (1.09) & 58.60, 96046 (0.45) & 39.26, 152113 (0.70) &
& \\

\midrule

\multirow{2}{*}{ML-KEM-768}
& 24  & -2.82, 384507 (16.02) & 69.88, 115392 (4.81) & 51.72, 196812 (8.20) &
\multirow{2}{*}{12168} & \multirow{2}{*}{+3488} \\

& 216 & -2.48, 387648 (1.80) & 69.79, 116972 (0.54) & 52.18, 197437 (0.91) &
& \\

\midrule

\multirow{2}{*}{ML-KEM-1024}
& 24  & -4.22, 615311 (25.64) & 74.60, 151781 (6.32) & 58.69, 260949 (10.87) &
\multirow{2}{*}{16488} & \multirow{2}{*}{+6176} \\

& 216 & -3.96, 618624 (2.86) & 74.33, 154431 (0.72) & 58.78, 262601 (1.22) &
& \\

\bottomrule
\end{tabular}%
}
\end{table*}

The key-generation penalty reflects construction of reusable state. For every parameter set and endpoint in Table~\ref{tab:recommended_profile}, the added key-generation cycles are smaller than the cycles saved by one subsequent encapsulation or decapsulation. This comparison does not include external distribution or persistent-storage costs. Both the storage requirement and the benefit grow with the module rank because the public matrix contains $k^2$ polynomials. At 216~MHz, the selected profile completes encapsulation in less than 1~ms for all three parameter sets.

\subsection{Clock-Scaling Evaluation}
\label{subsec:clock_frequency_effect}
The previous results compare optimization effects at the two edge-clock configurations. To separate cycle-level optimization from deployment latency, we additionally evaluate a clock sweep using the selected all-enabled, no-DTCM, no-parsed-PK profile; the complete 16-clock measurements are available in the artifact repository~\cite{MLKEMM7Repo2026}.

\begin{figure}[t]
\centering
\includegraphics[width=0.8\columnwidth]{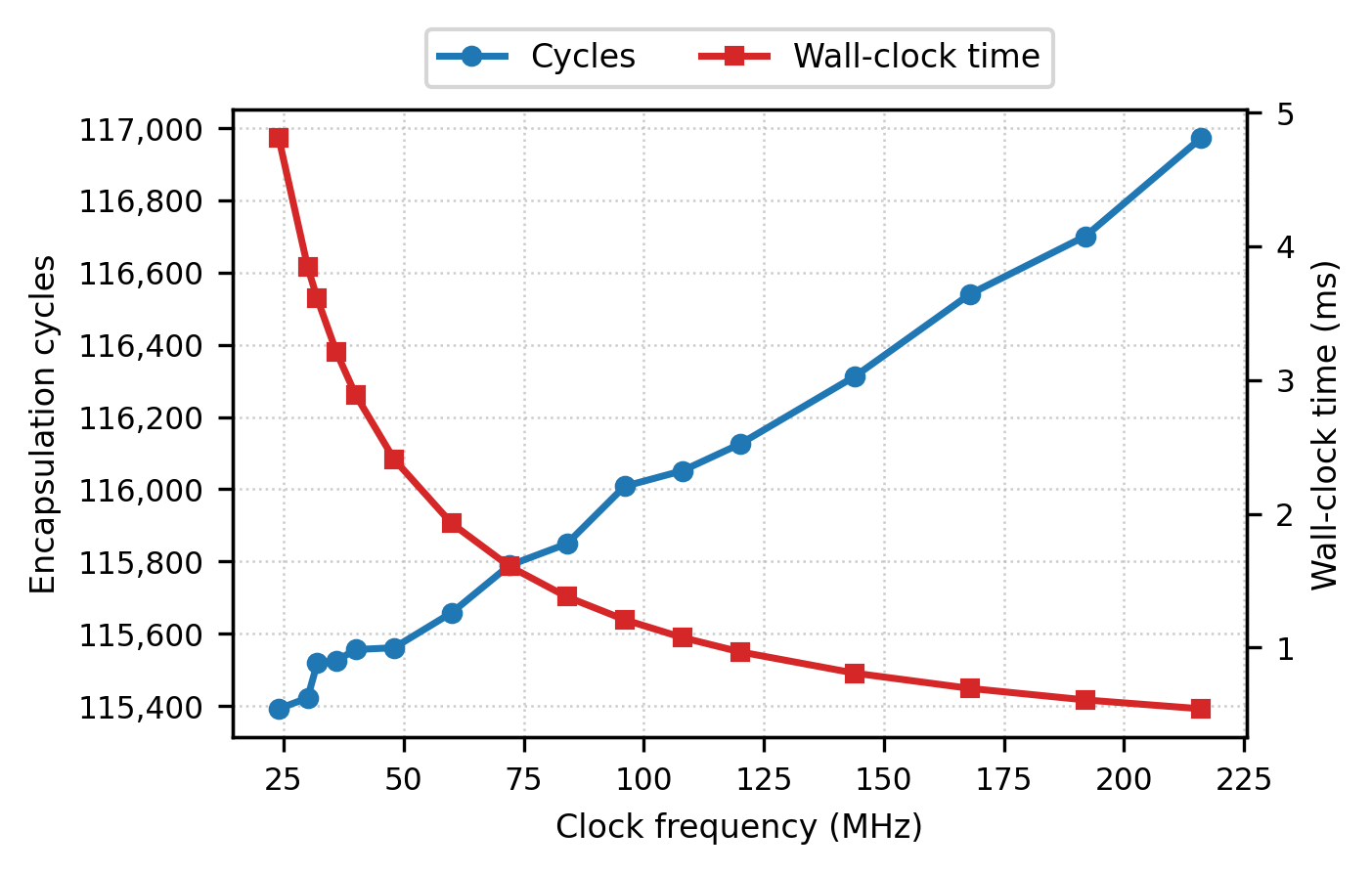}
\caption{Clock-frequency effect for ML-KEM-768 encapsulation using the selected all-enabled, no-DTCM, no-parsed-PK profile. Cycles use the left axis and wall-clock latency uses the right axis.}
\label{fig:clock_curve_768_encaps}
\end{figure}

Table~\ref{tab:clock_sweep_768_recommended} reports representative measurements for ML-KEM-768, while Figure~\ref{fig:clock_curve_768_encaps} illustrates the complete clock sweep for encapsulation.

\begin{table}[t]
\centering
\caption{Clock-frequency effect for ML-KEM-768 using the selected profile. Entries show average cycles and derived wall-clock time (ms) in parentheses.}
\label{tab:clock_sweep_768_recommended}
\scriptsize
\setlength{\tabcolsep}{3pt}
\begin{tabular}{rrrr}
\toprule
MHz & KeyGen & Encaps & Decaps \\
\midrule
24  & 384507 (16.02) & 115392 (4.81) & 196812 (8.20) \\
32  & 384610 (12.02) & 115518 (3.61) & 197097 (6.16) \\
48  & 384656 (8.01)  & 115561 (2.41) & 197031 (4.11) \\
60  & 384981 (6.42)  & 115658 (1.93) & 197061 (3.28) \\
72  & 385529 (5.36)  & 115789 (1.61) & 196979 (2.74) \\
96  & 385815 (4.02)  & 116008 (1.21) & 197097 (2.05) \\
120 & 386008 (3.22)  & 116125 (0.97) & 197089 (1.64) \\
144 & 386093 (2.68)  & 116313 (0.81) & 197161 (1.37) \\
168 & 386824 (2.30)  & 116541 (0.69) & 197271 (1.17) \\
216 & 387648 (1.80)  & 116972 (0.54) & 197437 (0.91) \\
\bottomrule
\end{tabular}
\end{table}

The measured cycle counts are not perfectly invariant across clock frequencies. For example, ML-KEM-768 key generation and encapsulation increase by approximately 0.8\% and 1.4\%, respectively, between 24~MHz and 216~MHz. This behavior is consistent with memory-system effects such as Flash wait states and cache-refill activity.

Nevertheless, the wall-clock impact of increased operating frequency dominates. In the representative ML-KEM-768 configuration, encapsulation latency decreases from 4.808~ms at 24~MHz to 0.542~ms at 216~MHz. Across all three parameter sets, increasing the clock frequency from 24~MHz to 216~MHz reduces wall-clock latency by approximately a factor of nine for all operations.

An important observation is that the latency improvement is not exactly proportional to clock frequency. Combined with the baseline behavior observed in Table~\ref{tab:baseline}, this illustrates that cycle-oriented benchmarking and latency-oriented deployment answer different questions. Cycle reductions characterize work at a fixed operating point, whereas clock scaling determines how quickly that work is completed on the target platform.

\subsection{Measurement Dispersion and Repeatability}
\label{subsec:measurement_repeatability}

Several effects in Tables~\ref{tab:isolated_effects} and~\ref{tab:combined_profiles} fall below 1\%, motivating a targeted check of measurement dispersion and repeatability. Rather than resampling the full matrix, we selected 11 configurations to probe memory-system knobs with small measured effects and to add limited checks of module rank and input-source variation. Eight compare the ML-KEM-768 baseline with Flash acceleration, DTCM constant placement, and ITCM code placement at 24 and 216~MHz, covering both frequency endpoints using the middle parameter set. ML-KEM-512 and ML-KEM-1024 baselines at 216~MHz check the other module ranks, while an ML-KEM-768 hardware-TRNG configuration at 216~MHz tests variability when inputs are refreshed.

Within each configuration run, the firmware computed each operation's standard deviation directly from the 100 timed iterations. Across 20 runs per configuration, this within-run standard deviation ranged 0.29--1.01\% of that run's mean with the fixed-seed generator and 0.31--0.67\% with the TRNG configuration, which puts the majority of the reported effects above the noise floor. 

Separately, to test the stability of the statistic used in our tables, each configuration was measured over 100 reboots with its binary held fixed, yielding another 100-iteration mean after each reboot. The fixed-seed generator replayed the same input sequence, whereas the TRNG supplied fresh inputs. Across reboots, the standard deviation of these means, normalized by their average, was at most 0.006\% for fixed-seed configurations and ranged from 0.050\% to 0.052\% with the TRNG. Thus, the reported 100-iteration averages are highly repeatable under the tested conditions. In particular, the ML-KEM-768 ITCM cycle savings were 0.40--0.45\% at 24~MHz and 1.28--1.45\% at 216~MHz, and every observed ITCM run mean was below every corresponding baseline run mean for all three operations at both endpoints.

\section{Deployment-Oriented Optimization Guidelines}
\label{sec:discussion_optimum_profiles}

No profile dominates every deployment objective. Key generation is penalized
by materializing public caches, encapsulation benefits from both cached $A$ and
$H(pk)$, and decapsulation benefits mainly from cached $A$. The appropriate
choice also depends on whether auxiliary public state can be retained, derived
once, or distributed with an authenticated association to the key.
Table~\ref{tab:deployment_profiles} translates the measurements into
operation- and constraint-oriented choices.

For example, consider a provisioned sensor that repeatedly establishes session
keys with a gateway whose authenticated ML-KEM public key remains stable. The
sensor can derive or receive the associated public state once and retain it for
subsequent encapsulations, while the gateway can retain keypair-associated state
prepared during key generation for later decapsulations. Both endpoints must
bind this state to the correct public key and invalidate it when that key is
rotated. This example illustrates amortization across a key lifecycle; the cost
of distributing or persistently storing the state is not measured here.

\begin{table}[t]
\centering
\caption{Evidence-supported profile choices for representative deployments. Measured minima compare the six composite profiles at 216~MHz.}
\label{tab:deployment_profiles}
\small
\setlength{\tabcolsep}{3pt}
\begin{tabular}{>{\raggedright\arraybackslash}p{0.23\textwidth}>{\raggedright\arraybackslash}p{0.25\textwidth}>{\raggedright\arraybackslash}p{0.39\textwidth}}
\toprule
Objective & Suggested profile & Rationale \\
\midrule
No auxiliary public state & NC-noD & Preserves standardized exchanged objects,
adds no public cache, and avoids ineffective DTCM placement. \\
Memory-limited encapsulation & $H(pk)$ cache with selected ITCM & Adds 32~B
and targets two measured encapsulation improvements. \\
Balanced complete flow & AE-noD-noPK & Minimizes the measured three-operation
sum for every parameter set while omitting DTCM and parsed-PK storage. \\
Minimum measured encapsulation & AE for all parameter sets & Lowest
encapsulation time among the six composite profiles. \\
Minimum measured decapsulation & AE for ML-KEM-512/768; AC for ML-KEM-1024 &
Lowest decapsulation time among the six composite profiles. \\
Minimum measured key generation & NC for ML-KEM-512; NC-noD for ML-KEM-768/1024 &
Avoids materializing public-reuse state. \\
\bottomrule
\end{tabular}
\end{table}

\FloatBarrier

NC-noD, AE-noD-noPK, AE, and AC are measured complete profiles. The $H(pk)$-plus-ITCM table entry and the cached-$A$-plus-ITCM candidate discussed below combine isolated public-reuse measurements with
execution-path-specific placement evidence; they are recommendations rather
than directly measured complete profiles. Such combinations should be
benchmarked as a whole before deployment because individually favorable
directions need not compose linearly.

\subsection{Communication-Constrained Deployments}

When no auxiliary public state can be distributed or retained, NC-noD is the
preferred measured default. It combines path-specific ITCM placement, hardware
TRNG, and Flash configuration without changing standardized exchanged objects,
and omits the DTCM constants that consumed 1792~B without measurable benefit in
isolation. Its cycle reductions are modest compared with public-data reuse, but
it avoids auxiliary-state communication, storage, and key-association
requirements.

For a key-generation-oriented endpoint at 216~MHz, NC is the measured minimum for ML-KEM-512, whereas NC-noD is the minimum for ML-KEM-768 and ML-KEM-1024. Both avoid materializing public-reuse state. Composite public-data-reuse profiles increase key-generation cycles by 1.08--6.27\% because they prepare state for later operations. This recommendation changes when a generated key is followed by
many encapsulations or decapsulations: in that lifecycle, the one-time
preparation cost may be amortized by subsequent savings.

\subsection{Memory-Constrained Deployments}

For an encapsulator with limited memory, cached $H(pk)$ is the clearest
storage--performance trade-off: its 32~B state reduces encapsulation cycles by
12.49--16.35\%. Pairing it with the selected ITCM placement for the
regenerated-$A$ path is an evidence-based candidate because ITCM independently
improves encapsulation without adding auxiliary public state. The complete
combination still requires measurement.

The cache must remain bound to the same public key and available for enough
operations to justify its preparation or distribution. By contrast, parsed
public-key caching requires 2--4~KiB for at most a 1.30\% isolated saving, and
DTCM constants consume 1792~B without corresponding cycle benefit. Cached $A$
provides much larger savings but requires 1.5--6~KiB, so it becomes attractive
only when the available memory and expected key-use lifetime support that
trade-off.

\subsection{Performance-Oriented Deployments}

For repeated use with sufficient memory, AE-noD-noPK is the balanced measured
profile. It retains the dominant cached-$A$ and $H(pk)$ savings and the other
system-level directions while omitting the low-value DTCM and parsed-public-key
storage. The fully enabled profile provides slightly larger reductions in some cases, but at substantially greater RAM and auxiliary-state cost.

Among the complete profiles at 216~MHz, AE gives the lowest measured decapsulation time for ML-KEM-512 and ML-KEM-768, while AC is lowest for ML-KEM-1024. A leaner decapsulation-oriented endpoint can instead investigate cached $A$ with captured-path ITCM: cached $A$ directly removes matrix expansion from the re-encryption path, whereas a separate $H(pk)$ cache does not help because decapsulation already obtains the digest from the standard secret key. This lean combination remains a candidate requiring complete-profile measurement.

Minimum wall-clock latency requires selecting among complete profiles at the
intended HCLK.\@ At 216~MHz, the balanced profile provides sub-millisecond
encapsulation for all parameter sets, but the fastest admissible profile still
depends on memory and state constraints. Higher frequency may increase power,
thermal load, and therefore battery or sustained-operation considerations.
Because this study measures neither power nor energy, it does not claim an
energy-optimal operating point or quantify device lifetime.

Overall, profile selection must account for operation mix, key lifetime,
memory, communication, and platform constraints rather than optimize a single
function invocation in isolation.

\section{Conclusion}
\label{sec:conclusion}

Starting from a SLOTHY-optimized Cortex-M7 implementation, this work evaluated
memory placement, Flash configuration, hardware randomness, operating
frequency, and deterministic public-data reuse across all three ML-KEM
parameter sets while preserving the standardized wire formats. Profiles without
auxiliary public state reduce cycles by up to approximately 2.5\%. The balanced
AE-noD-noPK profile reduces encapsulation and decapsulation cycles by up to
74.6\% and 58.8\%, respectively, at the cost of auxiliary public state and a
1.08--4.22\% key-generation increase. At 216~MHz, its encapsulation latency is
below 1~ms for every parameter set.

The results support continuing optimization beyond instruction-level kernels
to code and data placement, hardware integration, operating points, and
repeated work across API calls. On the evaluated platform, public-data reuse
provides the largest gains, while selective ITCM placement is the most
consistent improvement without auxiliary public state. The methodology and
selection criteria can guide evaluation on other embedded systems, but the
ranking of techniques and the resulting profiles remain platform, workload,
and lifecycle dependent.

\section*{Acknowledgements}
This project was funded in part by Mitacs Accelerate project IT49912 in collaboration with Quantegra Technologies Inc.

\bibliographystyle{splncs04}
\bibliography{OptM7}

\end{document}